\documentclass[11pt,preprint]{aastex}
\usepackage{amssymb}
\usepackage{epsfig}
\usepackage{natbib}
\usepackage{graphicx}
\usepackage{color}
\usepackage{tabularx}
\usepackage{epsfig}

\newcommand       \mum        {\,{\rm \mu m}}

\newcommand       \Ks           {{\rm K_{S}}}
\newcommand       \J            {{\rm J}}
\newcommand       \JJ            {{\rm J}}
\newcommand       \HH           {{\rm H}}

\newcommand       \K            {\,{\rm K}}
\newcommand       \simali       {\,{\sim}}
\newcommand       \magni        {\,{\rm mag}}

\newcommand       \pc           {\,{\rm pc}}

\newcommand       \Teff         {T_{\rm eff}}

\newcommand       \CJH       {C_{\rm JH}}
\newcommand       \CHK       {C_{\rm HK_S}}
\newcommand       \CJK       {C_{\rm JK_S}}
\newcommand       \CJHint       {C^0_{\rm JH}}
\newcommand       \CHKint       {C^0_{\rm HK_S}}
\newcommand       \CJKint       {C^0_{\rm JK_S}}
\newcommand       \EJH       {E_{\rm JH}}
\newcommand       \EHK       {E_{\rm HK_S}}
\newcommand       \EJK       {E_{\rm JK_S}}
\newcommand{\AV}{A_{\rm V}}

\newcommand{\AKs}{A_{\rm K_S}}

\newcommand{\AJ}{A_{\rm J}}
\newcommand{\RV}{R_{\rm V}}

\shorttitle{Universality of the Near-Infrared Extinction Law}
\shortauthors{Wang & Jiang}

\begin{document}

\title{
Universality of the Near-Infrared Extinction Law Based on the APOGEE Survey
     }

\author{Shu Wang\altaffilmark{1},
             B.W.~Jiang\altaffilmark{1}}
\altaffiltext{1}{Department of Astronomy,
                 Beijing Normal University,
                 Beijing 100875, China;
                 {\sf shuwang@mail.bnu.edu.cn,
                      bjiang@bnu.edu.cn
                 }
                 }

\begin{abstract}
Whether the near-infrared (NIR) extinction law is universal has been a long debated topic.  Based on the APOGEE H-band spectroscopic survey as a key project of SDSS-III, the intrinsic colors of a large number of giant stars are accurately determined from the stellar effective temperature. Taking this advantage and using a sample of 5942 K-type giants, the NIR extinction law is carefully re-visited. The color excess ratio $E(\JJ-\HH)/E(\JJ-\Ks)$, representative of the NIR extinction law, shows no dependence on the color excess when $E(\JJ-\Ks)$ changes from $\simali$0.3 to $\simali$4.0, which implies a universal NIR extinction law from diffuse to dense regions. The constant value of $E(\JJ-\HH)/E(\JJ-\Ks)$, 0.64, corresponds to a power law index of 1.95. The other two ratios, $E(\HH-\Ks)/E(\JJ-\Ks)$ and $E(\JJ-\HH)/E(\HH-\Ks)$, are 0.36 and 1.78 respectively. The results are consistent with the MRN dust size distribution.
\end{abstract}


\section{Introduction}

The early studies found that the near-infrared (NIR) ($0.9\mum < \lambda < 3\mum$) extinction follows a power law, $A_{\lambda}\propto{\lambda^{-\alpha}}$. Furthermore, the index was claimed to be constant as reviewed by \citet{Draine89}, since its value concentrates in a small range, such as $\alpha\approx$1.61 \citep{RL85},
1.70 \citep{Whittet88}, 1.75 \citep{Draine89}, 1.8 \citep{MW90, Whittet93}. This constancy points to a universal law in NIR even if the extinction law in the UV and visual wavebands changes significantly with the environment as indicated by the varying selective ratio R$_{\rm V}$ from $\sim$2.0 to $\sim$6.0 \citep{CCM89}.

This century presents a new view on the NIR extinction law. The power law index alpha has clearly taken different values and become systematically large, mostly $>$2.0 in comparison with previous 1.6-1.8. For example, Messineo et al.\ (2005) obtained a value of 1.9, supported by following measurements, 1.99 (Nishiyama et al.\ 2006), 2.07 (Strai\v zys \& Laugalys 2008), 2.64 (Gosling et al. 2009), 2.23 (Nishiyama et al.\ 2009), 2.14 (Stead \& Hoare 2009), 2.26 (Zasowski et al. 2009), 2.21 (Sch\"oedel et al. 2010), 2.11 (Fritz et al.\ 2011). One exception is 1.65 by \citet{Indebetouw05}.
When we investigated the infrared extinction towards five regions of the Coalsack nebula, the values of $\alpha$  were also larger than 2.1 in the translucent and dense clouds, while $\alpha$=1.73 in a diffuse region \citep{Wang13}.

There are a couple of uncertainties in determining the NIR extinction law. Using individual star, usually very luminous in infrared, suffers the uncertainty of the intrinsic infrared colors partly due to the spectral type and light variation of red giants.
On the other hand, statistical method that uses a group of stars of the presumably same spectral type relies completely on the photometric colors, which brings about the impurity because the photometric criteria unavoidably mix some objects with similar colors.
To make things more complicated, Stead \& Hoare (2009) argued that the power law index, very often taken as the measure of the NIR extinction law,  changes with the spectral energy distribution of the tracer that affects the effective wavelength ($\lambda_{\rm eff}$) of the filters.

With the data release DR10 of SDSS in 2013, the APOGEE survey provides the possibility of accurate determination of stellar intrinsic color index (CI) for numerous giant stars suitable for investigating the NIR extinction law. This work tries to take the advantages of both individual determination of intrinsic CIs and statistical method to study the NIR extinction law.

\section{Determination of the Intrinsic Colors of K-type Giants} \label{data}

\subsection{Data from the APOGEE Project}

Our study is based on the APOGEE project. APOGEE (The Apache Point Observatory Galaxy Evolution Experiment) is a large scale, high-resolution NIR spectroscopic survey of the Galactic stars, one of the four experiments in SDSS-III (Eisenstein et al.\ 2011). APOGEE targeted about 100,000 giant stars with {\it 2MASS} H magnitude down to 13$\magni$ with S$/$N$>$100 (Allende Prieto et al.\ 2008, Zasowski et al.\ 2013).
APOGEE measures the stellar parameters including effective temperature $\Teff$, surface gravity $\log g$ and metal abundance $Z$, to an accuracy of 150K in $\Teff$, 0.2 dex in $\log g$ and 0.1 dex in $Z$ (M\'esz\'aros et al.\ 2013). 

\subsection{The $\Teff$ - Color Relation of the APOGEE K-type Giants}

The NIR intrinsic colors of normal stars were defined by Johnson (1966) by adopting the average observed colors of stars within a distance of 100$\pc$ from the Sun. No interstellar extinction was corrected so that the derived intrinsic colors surpassed the real values. %
\citet{Ducati01} revised the Johnson result by expanding the sample to the Catalog of Infrared Observation, a database of over 396,000 infrared observations of $>$64,000 sources in the wavelength range from 1 to 1000$\mum$ (Gezari et al.\ 1999), and taking the bluest color as the intrinsic CI for a given spectral type. They naturally obtained systematically bluer colors than Johnson (1966). However, the uncertainty of CIs rises for the cold stars with the shrinking size of corresponding sample and the increase of scattering.

\begin{figure}[h!]
\centering
\vspace{-0.0in}
\includegraphics[angle=0,width=4.2in]{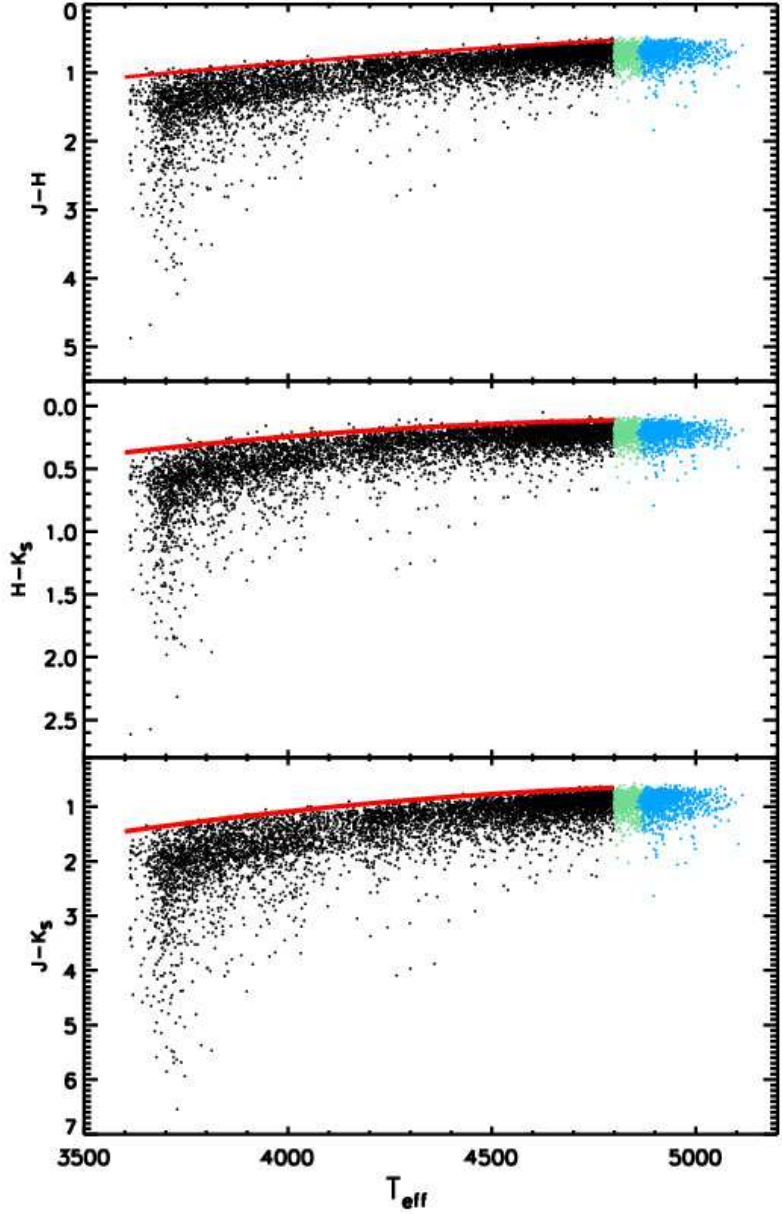}
\caption{\footnotesize
               \label{fig:intrinsic}
         The $\Teff$ versus observed colors $\CJH$, $\CHK$, and $\CJK$. Black dots are the selected 6074 K-type giants, blue dots the G-type giants and green dots the K-type giants possibly mixed with G-type giants. The red line denotes the fitting of the intrinsic colors.
          }
\end{figure}

We independently determine the intrinsic CIs of the APOGEE K-type giants which will be used to study the NIR extinction law. The APOGEE objects are originally from the {\it 2MASS} all-sky survey (Skrutskie et al.\ 2006) that measures the brightness in the $\JJ\HH\Ks$ bands and the observed CIs. By combining the stellar parameters from APOGEE and the observed CIs from {\it 2MASS}, the intrinsic CIs are derived through a method similar to Ducati et al.\ (2001), i.e. taking the blue envelop in the $\Teff$ vs. CI diagram.
Although this method can be applied to all the stars in the APOGEE sample, this work only deals with the APOGEE-nominated K-type giants to be used as the tracers of the NIR extinction law. Concentration on only K-type giants improves the reliability of the derived CIs thanks to a relatively narrow range of $\Teff$. For studying the extinction, the K-type giants already penetrate to deep extinction sightlines with the color excess (CE) $\EJK (\equiv E(\JJ-\Ks)) \simali 4.0$ (about $24\magni$ in $\AV$). In comparison, G-type giants trace shallower extinction (cf. blue dots in Figure~\ref{fig:intrinsic} with $\EJK$ mostly $<1.5$), and the stellar parameters of late M-type giants have not been accurately determined for the APOGEE survey partly due to the modeling difficulty.

As we can see in Figure~\ref{fig:intrinsic}, the $\Teff$---color diagram for the stars classified as giants by the APOGEE project, the ranges of  $\Teff$ of K-type and G-type giants have an overlap in $4800\K \le \Teff \le 5000\K$.  We set the upper limit of the K-type giants by $\Teff \leq 4800\K$, which coincides with the upper limit of K-type giant of \citet{BB88}. At the lower end, $\Teff$ agrees with the APOGEE catalog, $3500\K$, while in practice this occurs at $3600\K$, actually lower than the classical boundary of K-type giants. \citet{BB88} defined the lower boundary of K-type giants at $\sim3800\K$, and $\Teff=3600\K$ corresponds to a spectral type of M4. These possible early M-type giants are included because they still follow a well-defined $\Teff$---color relation and more importantly they trace large extinction. The surface gravity satisfies $\log g\leq3.0$ in accordance with the criterion for giants in the APOGEE catalog.
In addition, the tracing stars are required to have the photometric error $\leq0.05\magni$ in all the {\it 2MASS} bands. For obvious NIR extinction, we constrain within the Galactic plane by $\mid b \mid \le 5^{\circ}$. As the metallicity may affect the intrinsic color, $Z$ is limited to  $>-1.0$. The following summarizes the characteristics of the sample: (1) $3500\K\leq\Teff\leq4800\K$, (2) $\log g\leq3.0$, (3) $\sigma_{\J\HH\Ks}\leq0.05\magni$, (4) $\mid b \mid \le 5^{\circ}$ and (5) $Z>-1.0$. The final sample consists of 6074 APOGEE-nominated K-type giants. 

Figure~\ref{fig:intrinsic} shows the variation of the observed CIs with $\Teff$ of the sample giants (black dots) together with the G-type giants. The bluest stars in these diagrams of $\Teff$ versus observed color $\J-\HH$, $\HH-\Ks$, and $\J-\Ks$ (hereafter $\CJH$, $\CHK$ and $\CJK$) are considered to have neglectable extinction, and their observed colors are taken as the intrinsic ones. This idea is in principle the same as \citet{Ducati01}. But we have determined the bluest edge slightly differently.
As Figure~\ref{fig:intrinsic} shows, we choose the bluest stars  within a bin of $\delta \Teff=50\K$ at first and  then make a quadratic fitting to the bluest stars as following:
\begin{equation}
\CJHint= 4.37- 1.27\cdot(\frac{\Teff}{10^3\K})+0.098\cdot(\frac{\Teff}{10^3\K})^2 ~~,
\end{equation}
\begin{equation}
\CHKint= 3.35 - 1.29\cdot(\frac{\Teff}{10^3\K})+0.128\cdot(\frac{\Teff}{10^3\K})^2 ~~,
\end{equation}
\begin{equation}
\CJKint= 9.19 - 3.26\cdot(\frac{\Teff}{10^3\K})+0.309\cdot(\frac{\Teff}{10^3\K})^2 ~~,
\end{equation}
where $\CJHint, \CHKint$ and $\CJKint$ denote the intrinsic CIs ($\J-\HH)_0, (\HH-\Ks)_0$ and $(\J-\Ks)_0$.
The red lines in Figure~\ref{fig:intrinsic} show the fitting results.

Two factors are considered for the final adoption of the intrinsic CI at given $\Teff$. One is the measurement error of the CI. The selection of the photometric accuracy of 0.05$\magni$ at most brings about 0.1$\magni$ error in the CI, which not only causes the non-sharp edge but also makes the bluest stellar CI actually bluer than the true intrinsic CI. To compensate for this under-estimation, a redward shift of 0.02 and 0.03$\magni$ is added to $\CJHint$ and $\CHKint$ respectively as judged from visual inspection. The other factor is the consistency between the three CIs as only two are independent at a given $\Teff$. The intrinsic CIs at typical $\Teff$ are shown in Table~\ref{tab:color}.
The internal error of intrinsic colors $\mid\Delta\mid \equiv \mid\CJHint+\CHKint-\CJKint\mid$ are $\leq$ 0.02$\magni$.
Our NIR intrinsic CIs for classical K-type giants with $4000\K\le\Teff\le4800\K$ are $0.52\leq\CJHint\leq0.85$, $0.11\leq\CHKint\leq0.24$, and $0.65\leq\CJKint\leq1.09$, very consistent with previous results (e.g. \citealt{BB88}, \citealt{Wainscoat92}). The derived CIs at $3600\K$ however are about 0.1$\magni$ redder than, e.g. \citet{BB88}. Nonetheless, no modification is taken. This could be caused by the uncertainty in $\Teff$ that may happen in either the APOGEE project or previous work, and there is no sign that the red line in Figure \ref{fig:intrinsic} over-estimate the CI. Moreover, the CE will be calculated within this system so that the internal consistency should be retained.

\begin{table}[h!]
\begin{center}
\caption{\label{tab:color}
                The NIR intrinsic CIs at typical $\Teff$ for the selected sample giants}
\vspace{0.1in}
\begin{tabular}{ccccccccc}
\hline \hline
$\Teff (\K)$  & 3600 & 3800 & 4000 & 4200 & 4400 & 4600 & 4800\\
\hline
$\CJHint$  & 1.06 & 0.95 & 0.85 & 0.76 & 0.67 & 0.59 & 0.52 \\
$\CHKint$ & 0.37 & 0.30 & 0.24 & 0.20 & 0.16 & 0.13 & 0.11 \\
$\CJKint$  & 1.45 & 1.26 & 1.09 & 0.94 & 0.82 & 0.72 & 0.65 \\
\hline
$\mid\Delta\mid$ & 0.02 & 0.01 & 0.0 & 0.02 & 0.01 & 0.0 & 0.02 \\
\hline
\end{tabular}
\end{center}
\end{table}

\subsection{The Red Clump Stars}

The red clump (RC) (K2III) stars own the reputation with constant luminosity and small color scattering.
Their absolute magnitude is around
M$_{\rm K}= -1.61\magni$ (Alves 2000).
The intrinsic CI $\CJKint$ of RC stars centers around 0.75 (Wainscoat et al.\ 1992), or 0.65 \citep{GF14}.
Thus the RC stars are frequently used as the tracers of IR interstellar extinction.
The RC stars chosen by their clumping in the contour map of the $\Teff$---$\log g$ diagram have $4550\K \le \Teff \le 4800\K$ and $2.5 \le \log g \le 2.9$, consistent with $\Teff=4750\pm160\K$ and $\log g=2.41\pm0.26$ (Puzeras et al.\ 2010). Based on the $\Teff$---color relation derived above, the NIR intrinsic CIs of the RC stars are $0.52\leq\CJHint\leq0.61$, $0.11\leq\CHKint\leq0.14$, and $0.65\leq\CJKint\leq0.75$. It turns out that \citet{Wainscoat92} and \citet{GF14} gave the upper and lower limits respectively. The consistency with previous results on the RC stars reassures the correctness of our method of determining the intrinsic colors.

\section{The NIR Extinction Law}

To avoid the uncertainty in choice of the filter wavelength (effective or isophotal) when converting from a color excess ratio (CER) to a power law index, we take the CER as the measure of the NIR extinction law.
The CER depends much less on the filter wavelength than the power law index because the photometry is performed in wide-bands.

\subsection{Color Excess Ratio} \label{excess}

The calculation of CE is very straightforward from the difference between the intrinsic CI (derived from its dependence on $\Teff$) and the observed CI. 5942 stars are left in subsequent analysis after dropping 132 sources for their  negative CEs. In principle, the three CEs, $\EJH$, $\EHK$ and $\EJK$, are derived for every sample star and can be regarded as the indicator of the NIR extinction law individually. However, a statistical linear fitting between the CEs is adopted to alleviate significantly the uncertainty of individual measurement.

The intercept of linear fitting should be considered carefully while usually ignored. Physically, the CE, proportional to the total interstellar extinction, becomes zero when either of the other two CEs be zero. That means the intercept of the linear fitting between any two CEs should be zero ideally, or very close to zero with the uncertainty in the colors. In previous statistical studies of interstellar extinction, this constraint is rarely taken into account, very often because they calculated only the slope of the linear fitting of the observed CIs rather than the CER itself. 

\begin{figure}[h!]
\centering
\includegraphics[angle=0,width=3.8in]{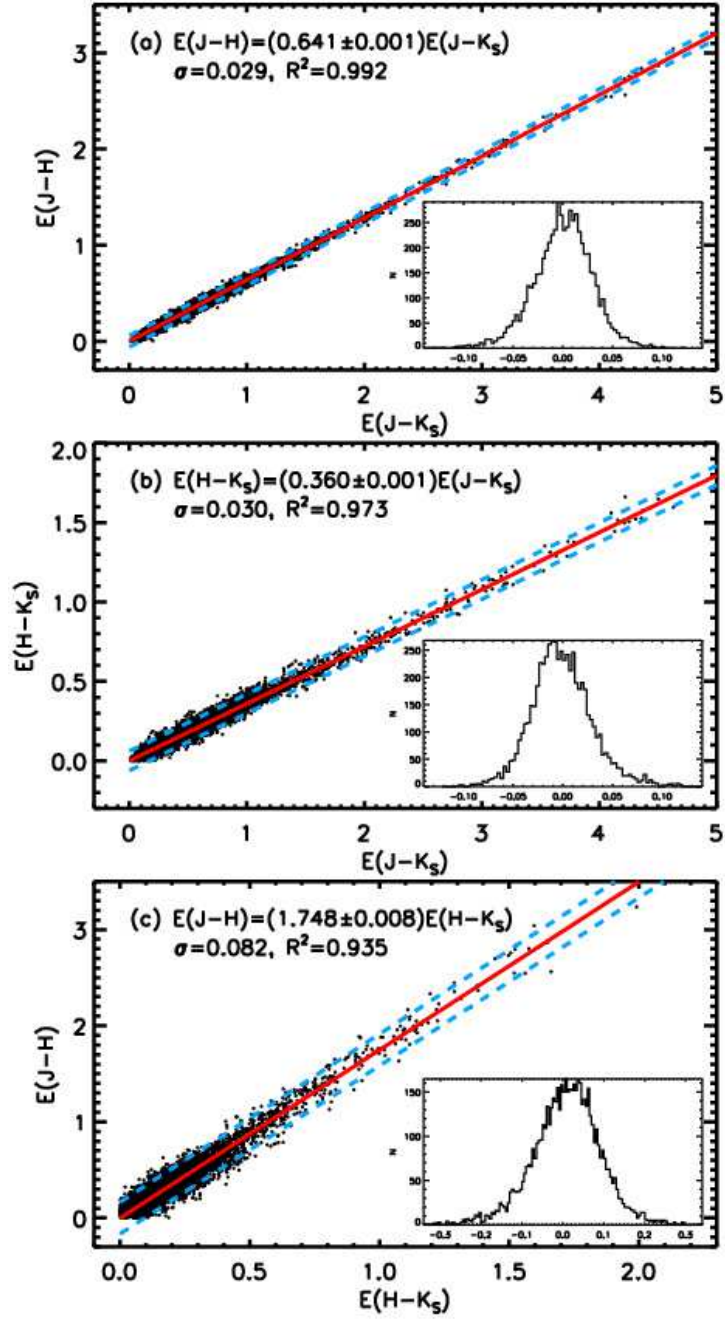}
\caption{\footnotesize
               \label{fig:excess}
          Relation between the NIR CEs of the 5942 K-type giants. Red line denotes the best linear fitting whose parameters are shown on. Inset shows the histogram of the residuals with blue dash line being the $2\sigma$ line.
               }
\end{figure}

The linear fitting results are displayed in Figure~\ref{fig:excess} as following: $\EJH/\EJK=0.641\pm0.001$, $\EHK/\EJK=0.360\pm0.001$, $\EJH/\EHK=1.748\pm0.008$, where the lines are forced to pass through (0,0). The histogram of the residuals is displayed as an inset, with the standard deviation of 0.029, 0.030 and 0.082 respectively.
The three ratios are independently calculated, with a very good internal consistency, as  $\EJH/\EJK$ and $\EHK/\EJK$ would yield $\EJH/\EHK=1.78$. Although any one of three ratios can be taken as the indicator of the NIR extinction law, the $\EJH/\EJK$ ratio is favored because of its large wavelength interval leading to the stability against uncertainty.  On the other hand, $\EJH/\EHK$ is very weak against the error, since $\EHK$ is only about third of $\EJK$. This weakness stands out particularly at small $\EHK$.
Nonetheless, this ratio was very often cited as the measure of the NIR extinction law possibly for its sensitivity to the variation of the NIR extinction power law index as shown in Table~\ref{tab:result}.

\subsection{Dependence of $\EJH/\EJK$ on $\EJK$}

This work takes the stars from all the fields surveyed by APOGEE with the Galactic longitude $0^{\circ} < l < 220^{\circ}$, and has no bias to any specific environment. Nontheless, the magnitude of CE represents in general the environment  because of its proportionality to the density of dust. Therefore, we investigate the variation of the CER $\EJH/\EJK$ along the CE $\EJK$ to clarify whether the NIR extinction law is universal.

\begin{figure}[h!]
\centering
\vspace{-0.0in}
\includegraphics[angle=0,width=5.5in]{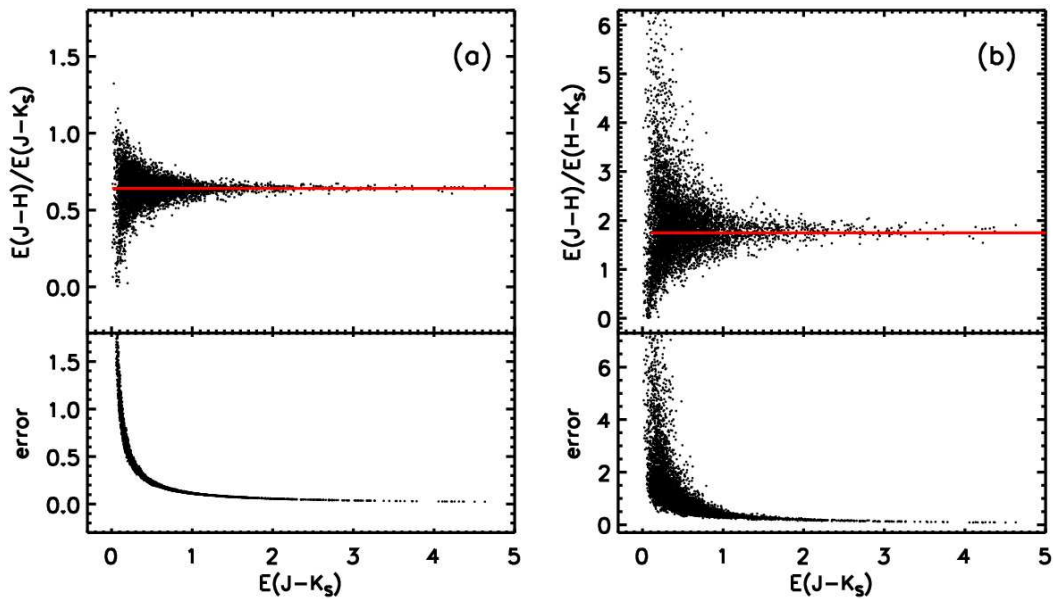}
\caption{\footnotesize
               \label{fig:ratio}
          The distribution of CER $\EJH/\EJK$ (a) and $\EJH/\EHK$ (b) with $\EJK$, with the red line as the CER derived from Figure~\ref{fig:excess} (a) and (c). Lower part is the error of the corresponding CER.
               }
\end{figure}

Figure~\ref{fig:ratio}(a) (upper) displays the values of $\EJH/\EJK$ from all the sample stars, with a red horizontal line highlighting the linear fitting result, i.e. $\EJH/\EJK=$0.641. It can be seen that all the stars are apparently around the red line. There is no clear systematic tendency towards either increasing or decreasing as $\EJK$ changes from very small value representative of diffuse ISM to $\EJK\sim5\magni$ equivalent to a visual extinction of $\sim$30$\magni$ attainable only through dense regions. A correlation analysis results in a Pearson correlation coefficient of 0.03 indicative of no relation between $\EJH/\EJK$ and $\EJK$.  On the other hand, the dispersion in $\EJH/\EJK$ is apparent and presents an increasing tendency when $\EJK$ gets small. Whether this dispersion is genuine needs to take into account the error.

The error of the $\EJH/\EJK$ values comes from a few contributors. The primordial errors originate from that of $\Teff$ and photometry. The average error of the APOGEE stellar parameter $\Teff$ is $\simali100\K$. Using it to derive the NIR intrinsic colors by equation (1) brings about an average error of 0.05, 0.02, 0.07$\magni$ respectively for $\CJHint, \CHKint, \CJKint$. With constraint on the photometric quality of selected stars $\sigma_{\J\HH\Ks}\leq0.05\magni$, the average photometric error is $\simali0.02$, consequently the average error of observed CI is $\simali0.04$ ($<1\%$ stars have the observed color error $\simali 0.1$).
Combining the photometric error in the $\J\HH\Ks$ bands and the error in the intrinsic colors, the uncertainties of the CEs are $(\EJH)$$_{\rm err} \simali0.09$, $(\EHK)$$_{\rm err} \simali0.06$, and $(\EJK)$$_{\rm err} \simali0.11$. Given these errors in the CEs, the error of CER $(\EJH/\EJK)$$_{\rm err}$ which depends on both $\EJH$ and $\EJK$ can be calculated under the error propagation theory. The error calculated by this method is displayed in Figure~\ref{fig:ratio}(a) (lower) for all the sources.
The error rises rapidly as $\EJK$ decreases. For a given $\EJH/\EJK$=0.641, the error $(\EJH/\EJK)$$_{\rm err}$ at $\EJK$=0.3 is 10 times larger than at $\EJK$=3, specifically, from 0.38 to 0.038. At $\EJK$=0.1, the error reaches 1.14. The error amplitude and its tendency both agree very well with the dispersion of $\EJH/\EJK$ in Figure~\ref{fig:ratio}(a) (upper). Therefore, the dispersion can be fully explained by the error.

The case for $\EJH/\EHK$ is shown in Figure~\ref{fig:ratio}(b). The error is about three times that of $\EJH/\EJK$, analogous to the dispersion. A correlation analysis yields a Pearson correlation coefficient of 0.05 indicative also of no relation between $\EJH/\EHK$ and $\EJK$. This result confirms non-variation of the NIR extinction law.

Because there is no apparent tendency with increasing reddening and with the dispersion accountable by the error of CER, we conclude no variation of the ratio $\EJH/\EJK$ with $\EJK$, i.e. the extinction law in the NIR $\J\HH\Ks$ bands is universal from diffuse to dense interstellar clouds.

\section{Discussion}

Various parameters delineate the NIR extinction law. Often used are the power law index $\alpha$, the three CERs or the extinction normalized to the K band. For the convenience of comparison, we calculated the corresponding $\alpha$ and $\AJ/\AKs$ from the CERs. The values of $\alpha$ are 1.95$^{+0.02}_{-0.01}$, 1.95$^{+0.02}_{-0.02}$, 1.88$^{+0.02}_{-0.02}$ derived from $\EJH/\EJK$, $\EHK/\EJK$ and $\EJH/\EHK$ respectively when adopting $\lambda_{\rm eff}$ of $\J\HH\Ks$ bands at 1.25, 1.65 and 2.15$\mum$, yielding $\AJ/\AKs$=2.88 at $\alpha$=1.95\footnote{If the standard deviations of the residual of the linear fitting are taken as the uncertainty of the CERs, the derived $\alpha$ with errors becomes 1.95$^{+0.47}_{-0.46}$, 1.95$^{+0.49}_{-0.47}$, 1.88$^{+0.17}_{-0.18}$ from $\EJH/\EJK$, $\EHK/\EJK$ and $\EJH/\EHK$.}.

In addition, the unavailable parameters are calculated for previous works with the provided information on CER or the power law index and the alpha value is re-calculated from CER, as shown in Table~\ref{tab:result} where the boldface denotes the values from the reference and the normal font denotes the values converted by ourselves using $\lambda_{\rm eff}$ of {\it 2MASS}.  It can be seen that our result agrees with almost the average of previous works for different sightlines towards diversified environments. The $\alpha$ value of this work, 1.95, is larger than the widely derived values 1.6-1.8 in 1980s, meanwhile agrees with the works of this century. This is partly caused by the difference of $\lambda_{\rm eff}$ of K band of the Johnson system and the {\it 2MASS} system $\Ks$. For a given $\AJ/\AKs$(=2.88), the {\it 2MASS} system would yield a larger $\alpha$, e.g. $\alpha$ decreases from 1.95 to 1.84 when $\lambda_{\rm eff}$ of K band shifts from 2.15$\mum$ for {\it 2MASS} to 2.22$\mum$ for the Johnson system. The other possible reason comes from the use of $\lambda_{\rm eff}$ instead of the isophotal wavelength ($\lambda_{\rm iso}$). As for the {\it 2MASS} system, adopting $\lambda_{\rm iso}$, 1.24, 1.66 and 2.16$\mum$ for the $\J\HH\Ks$ band, would yield $\alpha=1.65$ and $\AJ/\AKs$=2.51 for given $\EJH/\EJK$=0.64, consistent with the smaller values of \citet{Indebetouw05} who used $\lambda_{\rm iso}$.

\begin{table}[h!]
\small
\begin{center}
\caption{\label{tab:result}
                Summary of the results and comparison with previous works}
\vspace{0.1in}
\begin{tabular}{ccccccccc}
\hline \hline
  Works & $\frac{\EJH}{\EJK}$ & $\frac{\EHK}{\EJK}$ & $\frac{\EJH}{\EHK}$ & $\AJ/\AKs$
  & $\alpha$ & Environment \\
\hline
\textbf{This work} & \textbf{0.64} & \textbf{0.36} & \textbf{1.78} & \textbf{2.88} & \textbf{1.95} & \textbf{average}\\
Landini et al.\ (1984) & 0.62(\textbf{0.62}) & 0.38 & 1.71 & 2.42 & 1.63(\textbf{1.85}) & G333.6-0.2 (HII) \\
Martin \& Whittet 1990 & 0.63 & 0.37 & 1.71  & 2.65 & \textbf{1.8} & diffuse ISM\\
Racca  et al.\ (2002) & 0.68 & 0.32 & \textbf{2.08} & 3.92 & 2.52 & Coalsack globule 2\\
Indebetouw et al.\ (2005) & 0.64 & 0.36(\textbf{0.36}) & 1.78 & 2.86(\textbf{2.50}) & 1.94(\textbf{1.65}) & $l=42^\circ$ and 284$^\circ$\\
Naoi et al.\ (2006) & 0.62 & 0.38 & \textbf{1.66} & 2.50 & 1.69 & $\rho$ Oph, Chamaeleon \\
Nishiyama et al.\ (2006) & 0.64(\textbf{0.58}) & 0.36(\textbf{0.34}) & 1.80(\textbf{1.72}) & 2.94 & \textbf{1.99}\tablenotemark{\ast} & Galactic center\\
Stead \& Hoare (2009) & 0.65 & 0.35 & 1.88 & 3.19 & \textbf{2.14} & $27^\circ< l < 100^\circ$\\
Wang et al.\ (2013) & 0.65 & 0.35 & 1.86(\textbf{1.86}) & 3.14(\textbf{3.14}) & 2.11(\textbf{2.10}) & Coalsack\\
WD01 $\RV$=3.1 & 0.62 & 0.38 & 1.63 & \textbf{2.40} & 1.62 & diffuse\\
WD01 $\RV$=5.5 & 0.62 & 0.38 & 1.62 & \textbf{2.38} & 1.60 & dense \\
\hline
Average of previous works & 0.64 & 0.36 & 1.77 & 2.84 & 1.90 & diversified \\
\hline
\hline
\end{tabular}
\tablenotetext{\ast}{The alpha values derived from $\EJH/\EJK$, $\EHK/\EJK$ and $\EJH/\EHK$ are 1.01, 2.26 and 1.82 respectively when using the {\it 2MASS} $\lambda_{\rm eff}$.}
\end{center}
\end{table}

We investigate one more possible reason for the discrepancy, the effect of metallicity. Gao et al.\ (2013) obtained a value of $\EJH/\EHK$=1.25 for the LMC NIR extinction which agrees well with previous studies but is significantly lower than the Galactic value, meanwhile their $\EJH/\EJK$=0.64 coincides very well with present work. A metal-poor sample of 735 giants in the whole APOGEE sky is selected under the same criteria
but with  $Z <$ -1.0. Most of them are located in the halo as expected and with low extinction ($\EJK<1$). Using the same method as for the non-metal-poor giants, we obtained the NIR CERs  $\EJH/\EJK$=0.73, $\EHK/\EJK$=0.36 and thus $\EJH/\EHK$=2.03, which exhibits some difference from the non-metal-poor sample. But the tendency is opposite to the work of Gao et al.\ (2013). Due to mainly the low extinction and also the small number of stars, these results are quite uncertain. On the other hand, the LMC is not so poor as the sample stars. Whether and how the metallicity affects the NIR extinction law needs further investigation.

The \citet{WD01} (WD01) dust model produces invariant NIR extinction law when $\RV$ changes from 3.1 to 5.5, corresponding to the power law index from 1.62 to 1.60 as shown in Table~\ref{tab:result}. This can explain the universality of the NIR extinction law even though the change of dust size distribution leads to apparent variation in the optical extinction law. Their results are consistent with ours when using the standard deviations as uncertainties for alpha.  
On the other hand, if assuming the dust size distribution conforms the classical power law with an index of 3.5 \citep{MRN},  our model calculation \citep{Wang13} yields $\EJH/\EJK$=0.65 when $a_{\max}$, the maximum cutoff radius of the spherical dust grains, occurs at 0.25$\mum$. This means the dust size distribution of the MRN model better matches our result.

\section{Summary}

Based on the NIR spectroscopic survey project APOGEE, a sample of giant stars consisted of mainly K-type and some early M-type stars is selected. The relations between the effective temperature and three NIR intrinsic colors are constructed by fitting the bluest colors with a quadratic line.
When the extinction changes from small to very large value, the CERs, indicator of the NIR extinction law, shows no apparent variation.
The constant CERs are $\EJH/\EJK$=0.64, $\EHK/\EJK$=0.36, and $\EJH/\EHK$=1.78. The $\EJH/\EJK$=0.64 is converted to a power law index of 1.95 given $\lambda_{\rm eff}$ of {\it 2MASS}. This result is consistent with the MRN dust size distribution.

\acknowledgments{We thank Xiaodian Chen, Jian Gao, Mingjie Jian and Aigen Li for very helpful discussions, and the referee for very suggestive comments. This work, using the database from SDSS-III, is supported through China's grants NSFC 11373015 and 973 Program 2014CB845702.
}

\end{document}